\documentstyle[11pt,aaspp4]{article}

\slugcomment{submitted to {\it the Astrophysical Journal, Letters}}

\lefthead{kato and Hachisu}
\righthead{Helium Shell Flashes toward SN Ia Explosions}

\def\gtsim {>\kern-1.2em\lower1.1ex\hbox{$\sim$}~~}   
\def\ltsim {<\kern-1.2em\lower1.1ex\hbox{$\sim$}~~}   

\begin{document}

\title{A NEW ESTIMATION OF MASS ACCUMULATION EFFICIENCY 
IN HELIUM SHELL FLASHES TOWARD TYPE Ia SUPERNOVA EXPLOSIONS}

\author{Mariko Kato}
\affil{Department of Astronomy, Keio University, 
Hiyoshi, Kouhoku-ku, Yokohama 223-8521, Japan \\ e-mail: 
mariko@educ.cc.keio.ac.jp}

\begin{center}
and 
\end{center}

\author{Izumi Hachisu}
\affil{Department of Earth Science and Astronomy, 
College of Arts and Sciences, University of Tokyo,
Komaba, Meguro-ku, Tokyo 153-8902, Japan \\ e-mail: 
hachisu@chianti.c.u-tokyo.ac.jp}




\begin{abstract}
We have calculated the mass accumulation efficiency during helium shell 
flashes to examine whether or not a carbon-oxygen white dwarf (C+O WD)
grows up to the Chandrasekhar mass limit to ignite a Type Ia supernova
(SN Ia) explosion.  It has been frequently argued that 
luminous super-soft X-ray sources (SSSs) and symbiotic stars are 
progenitors of SNe Ia.  In such systems, 
a C+O WD accretes hydrogen-rich matter from a companion and burns 
hydrogen steadily on its surface.  The WD develops a helium layer 
underneath the hydrogen-rich envelope and undergoes periodic helium 
shell flashes. Using OPAL opacity, 
we have reanalyzed a full cycle of helium shell flashes on 
a $1.3 M_\odot$ C+O WD and confirmed that the helium
envelope of the WD expands to blow a strong wind. 
A part of the accumulated matter is lost by the wind.
\par
     The mass accumulation efficiency in helium shell 
flashes is estimated as
$
\eta_{\rm He} = -0.175 \left( \log \dot M + 5.35 \right)^2 + 1.05,
$
for $ -7.3 < \log \dot M < -5.9$, and $\eta_{\rm He}=1$ 
for $ -5.9 \le \log \dot M \ltsim -5$,
where the mass accretion rate $\dot M$ 
is in units of $M_\odot ~{\rm yr}^{-1}$.
In relatively high mass accretion rates 
as expected in recent SN Ia progenitor models, 
the mass accumulation efficiency is large 
enough for C+O WDs to grow to the Chandrasekhar mass,
i.e., $\eta_{\rm He} = 0.9$ for $\log \dot M = -6.3$, 
and $\eta_{\rm He}=0.57$ for $\log \dot M = -7.0$.
The wind velocity ($\sim 1000$ km s$^{-1}$) is much faster
than the orbital velocity of the binary ($\ltsim 300$ km s$^{-1}$) 
and therefore, the wind cannot be accelerated further by the companion's 
motion.  We suggest 
observational counterparts of helium shell flashes 
in relation to long term variations in super-soft X-ray fluxes
of SSSs and symbiotic stars.

\end{abstract}




%

\section{INTRODUCTION}
Type Ia supernovae (SNe Ia) are widely believed to be a thermonuclear 
explosion event of a white dwarf (WD), although the immediate progenitors 
have not been identified yet.  Among many evolutionary paths toward SNe Ia 
proposed so far, arguments are now focused on 
the Chandrasekhar (Ch) mass model vs. the sub-Chandrasekhar (sub-Ch) 
mass model, and the single degenerate (SD) model vs. the double 
degenerate (DD) model (e.g., \cite{bra95} for a review 
of SN Ia progenitors). Of these models, the SD \& Ch model is the most 
promising one both from the observational and theoretical points of view 
(e.g., \cite{nom94} for a review; \cite{kob98}). 

One example of the SD \& Ch models is 
luminous super-soft X-ray sources (SSSs) in which the WD accretes 
hydrogen-rich material at relatively high rates ($\dot M \gtsim 1 
\times 10^{-7} M_\odot$ yr$^{-1}$) (van den Heuvel et al, 1992).  
The WD accretes almost pure helium ($Y=0.98$) as a results of stable
hydrogen shell-burning and experience helium shell flashes.  
In a strong flash, the envelope 
expands greatly (e.g., \cite{ibe82}) 
and may suffer wind mass-loss or a Roche lobe overflow 
(e.g., \cite{ksh89}). 
If a large part of the helium layer will be lost by the wind,
the WD hardly grows in mass to the Chandrasekhar mass (\cite{cit98}). 
Thus, it is essentially important to estimate the efficiency of 
the net mass accumulation onto the C+O core.  Kato, Saio, \& Hachisu
(1989, hereafter KSH89) has first estimated 
the mass accumulation efficiency in 
helium shell flashes.  They used the old opacity, however, and 
it should be recalculated with the new opacity.  

A new evolutionary path of the SD \& Ch scenario has been proposed by
Hachisu, Kato, \& Nomoto (1996; hereafter HKN96), 
in which a WD accretes hydrogen-rich matter from a lobe-filling 
red-giant (WD+RG system).  In this model, optically thick 
winds blow from WDs, which stabilize the mass transfer 
even if the donor has a deep convective envelope.
The WD steadily accretes hydrogen-rich matter from the companion. 
Its mass eventually reaches the Chandrasekhar mass limit and the WD 
explodes as an SN Ia. 
Hachisu, Kato, \& Nomoto (1999a, hereafter HKN99) show this
WD+RG system is one of the main channels to SNe Ia.  
Li \& van den Heuvel (1997) extended HKN96's model to another type of 
binary systems consisting of a mass-accreting WD and a lobe-filling,
more massive, somewhat evolved main-sequence star (WD+MS system). 
Hachisu \& Kato (1999) further 
proposed an evolutional path to binaries consisting of a WD and 
a somewhat evolved MS star characterized 
by helium-rich accretion as observed in U Sco. 
This is another main channel to SNe Ia as shown by
Hachisu, Kato, Nomoto, \& Umeda (1999b, hereafter HKNU99). 

All of these models will experience helium shell flashes 
followed by optically thick winds during its way to an SN Ia. 
If most of the helium matter is 
blown off in the wind, the WD cannot grow to the Chandrasekhar mass
($M_{\rm Ch}= 1.38 M_\odot$).
Therefore, once again it is essentially important to evaluate 
the efficiency of mass accumulation during the wind phase of 
helium shell flashes
in order to reach the definite conclusion of SN Ia progenitors.

Cassisi, Iben \& Tornamb\'e 
(1998, hereafter CIT98) criticized HKN96's model and concluded
that the white dwarf mass hardly 
increases to $M_{\rm Ch}$ because of heavy mass loss owing to a  
Roche lobe overflow during helium shell flashes. 
However, once the optically thick wind occurs during helium shell flashes, 
the wind velocity is as fast as $\sim 1000$ km s$^{-1}$, 
much faster than the orbital velocity ($\ltsim 300$ km s$^{-1}$) so that 
the companion motion hardly affects the mass ejection.  
The wind matter quickly goes away from the system without any 
interaction to the orbital motion.   Therefore their argument has no 
physical meaning when a strong wind occurs.

In this {\it Letter}, we have followed a full cycle of helium shell 
flashes including the mass loss owing to optically thick winds 
in order to obtain the mass accumulation efficiency of accreting WDs. 
We have used the same techniques as in KSH89 but adopted 
OPAL opacity that has a strong peak around 
$\log T ~(\mbox{K}) \sim 5.2$, which drives 
a strong wind as shown by Kato \& Hachisu
(1994).  In \S 2, we describe our methods, assumptions, and main results 
of helium shell flashes on a $1.3 M_\odot$ WD. 
Discussions follow in \S 3.

\section{MASS LOSS DURING HELIUM SHELL FLASHES}
Kato, Saio, \& Hachisu (1989, KSH89) adopted 
two different approaches to follow  
a full cycle of helium shell flashes: time-dependent hydrostatic 
calculation and mimicking time-dependent behavior by a sequence 
of steady-state wind solutions.  They combined these  
two methods, because the time-dependent Henyey code failed 
when the envelope expands greatly while the steady-state approach works 
well.  We have used the same basic approaches as 
in KSH89 to follow a full cycle of helium shell flashes 
on a $1.3 M_\odot$ WD.
The main difference is in the wind solutions, which we replaced 
by the new solutions recalculated 
with the updated OPAL opacity (\cite{ir96}).

The ignition models are unchanged because OPAL opacity gives essentially 
the same values as the old opacity does at higher temperatures of
$\log T ~(\mbox{K}) > 6$. 
We have assumed the same white dwarf radius of $1.3 M_\odot$, i.e.,
$\log R ~({\rm cm})= 8.513$, 
and the envelope chemical composition of $Y=0.48$, 
$C+O=0.5$ and $Z=0.02$ for helium, carbon and oxygen, and heavy 
elements (including C+O in solar ratio) by weight, respectively. 
Such a small helium content is expected as follows: 
in the early stage of a shell flash, the convection spreads
over the entire envelope, then descends and disappears 
before the photosphere reaches the maximum expansion. 
The ashes of helium burning, i.e., carbon and oxygen, is mixed into 
the entire envelope to reduce the helium content by about a 
half on average from the initial value of $Y=0.98$.  
Thus, we have assumed the uniform composition with 
$Y=0.48$ in the quasi-evolutionary sequence. 
The convective energy transport is included in both the static and the  
wind solutions by the mixing length theory with the mixing-length parameter 
of 1.5. The convection is, however, inefficient in the wind solutions,
because a wind is too fast (supersonic) compared with 
the convective motion (subsonic).  (See \cite{kat94} for details.)

Figure 1 depicts evolutionary tracks of one cycle of helium shell flashes.
After the onset of a shell flash, the helium layer expands 
to reach the maximum expansion of the photosphere, 
the radius of which depends
on the ignition mass $\Delta M_{\rm ig}$ of the helium layer.
We call this {\it rising phase}.  
In the rising phase, the optically thick
wind occurs when the photospheric temperature 
decreases to $\log T_{\rm ph} \sim 5.45$.
After the maximum expansion, the photospheric radius decreases with 
the envelope mass and the
star moves blueward in the HR diagram. After the wind stops, the envelope 
mass further decreases owing to nuclear burning and the star moves 
downward to come back to the original position. 
We call this {\it decay phase}.

For the rising phase, evolutionary tracks 
are approximated by a sequence of static
solutions with a constant envelope mass 
until the optically thick wind occurs
(see KSH89 for details).  Three tracks have been calculated with 
(a) $\Delta M_{\rm ig}= 4.8 \times 10^{-5} M_\odot$, 
(b) $\Delta M_{\rm ig}= 8.7 \times 10^{-5} M_\odot$, and 
(c) $\Delta M_{\rm ig}= 3.5 \times 10^{-4} M_\odot$. 
These ignition masses are realized 
for the helium accretion rates of  
(a) $\dot M_{\rm acc}= 1.7 \times 10^{-6} M_\odot$ yr$^{-1}$, 
(b) $\dot M_{\rm acc}= 6.7 \times 10^{-7} M_\odot$ yr$^{-1}$, and
(c) $\dot M_{\rm acc}= 1\times 10^{-7}M_\odot$ yr$^{-1}$,
respectively. 
The helium accretion rates of (a) and (b) correspond to the largest rates of 
hydrogen steady shell-burning,  
$\dot M_{\rm acc} \sim L_{\rm Edd}/X\varepsilon_H$,
in the envelope with $X=0.35$ and $X=0.7$, respectively, where 
 $L_{\rm Edd} \propto 1/(1+X)$ is the Eddington luminosity 
for electron scattering and $\varepsilon_H$ is 
the nuclear energy release of hydrogen-burning per unit mass.
Thus $\dot M_{\rm acc}$ 
is proportional to $1/X(1+X)$ for the hydrogen content $X$.  
Such a low hydrogen content of $X=0.35$ (or $Y=0.63$) is expected 
in the helium-rich matter accretion in recurrent nova U Sco (\cite{hak99}).

The evolutionary track 
for (a) $\Delta M_{\rm ig}= 4.8 \times 10^{-5} M_\odot$ 
is very close to the curve for the decay phase (thick curve)  
and omitted in Figure 1.  
No optically thick winds occur in this case because
the photospheric temperature does not decrease 
to $\log T_{\rm ph} \sim 5.45$.
At the maximum expansion,
the photospheric temperature reaches $\log T_{\rm ph}= 5.83$ 
(indicated by arrow a), 
and the radius $R_{\rm ph} = 0.022 R_\odot$.

In the case of (b) $\Delta M_{\rm ig}= 8.7 \times 10^{-5} M_\odot$, 
the wind mass-loss begins at $\log T_{\rm ph}= 5.45$. 
The maximum expansion 
is reached at $R_{\rm ph} \le 0.70 R_\odot$ and
$\log T_{\rm ph} \ge 5.05$ (indicated by arrow b), 
where the equality stands 
when we neglect the wind mass loss in the rising phase. 
The star moves blueward along the thick curve in the decay 
phase.  The envelope mass is decreasing owing to wind 
and nuclear burning.  The optically thick wind ceases at 
$\log T_{\rm ph}= 5.45$.  
In the third case of (c) $\Delta M_{\rm ig}= 3.5 \times 10^{-4} M_\odot$, 
the wind mass loss begins at $\log T_{\rm ph}= 5.44$, and the star 
reaches the maximum 
expansion at $R_{\rm ph} \le 9.1 R_\odot$ and $\log T_{\rm ph} \ge 4.45$  
(indicated by arrow c).

These tracks for three different envelope 
masses merge into one track after the maximum expansion, 
because the envelope 
reaches a thermal equilibrium in the decay phase.
The more massive the ignition mass is, 
the more redward the star moves in the HR diagram,
that is,
the star is traveling a longer track to a lower temperature. 
If the ignition mass is as large as 
$\Delta M_{\rm ig} \sim 7.7 \times 10^{-4} M_\odot$, 
the star reaches $\log T_{\rm ph} \ge 4.07$.
This envelope mass, $\Delta M_{\rm ig} = 7.7 \times 10^{-4} M_\odot$ 
corresponds to the 
ignition mass at 
$\dot M_{\rm acc} = 4.5 \times 10^{-8} M_\odot~$yr$^{-1}$.

Figure 2 shows the photospheric temperature $T_{\rm ph}$, 
the photospheric radius $R_{\rm ph}$,
and the photospheric wind velocity $V_{\rm ph}$ 
in the decay phase of helium shell flashes. 
The arrows indicate the same three ignition masses as in Figure 1.
At the maximum expansion, the star reaches these position
and then moves blueward as the envelope mass decreases 
owing to wind mass-loss and nuclear burning. 
The wind mass-loss stops at $\Delta M= 5.0 \times 10^{-5} M_\odot$,
marked by a filled circle.
The photospheric velocity is as large as $\sim 1000$ km s$^{-1}$,
which is much faster than the previous results 
($\sim 300$ km s$^{-1}$ in KSH89),
because of a strong acceleration 
driven by the large peak of OPAL opacity. 

The newly obtained wind velocities are much faster than 
the orbital velocities of SN Ia progenitors of 
the WD+MS or WD+RG systems ($\ltsim 300$ km s$^{-1}$).
Therefore, we can neglect effects of the binary motion 
on the mass ejection during the common envelope phase, because 
the envelope matter quickly goes away with little interaction 
with the orbital motion. This implies that neither a Roche lobe overflow 
nor a common envelope action plays an important role on mass ejection.
 
Figure 3 shows the total mass decreasing rate of the envelope 
$\log | d(\Delta M)/dt| $, which is the sum of the wind mass 
loss rate $\dot M_{\rm wind}$ and the mass decreasing rate 
owing to nuclear burning $\dot M_{\rm nuc}$ against 
the envelope mass $\log \Delta M$ 
for the decay phase of helium shell flashes. The processed 
matter accumulates at a rate higher than the mass ejection rate
for the region of $\log \Delta M < -3.74$,
and then a large part of the envelope matter 
accumulates on the white dwarf. 
These results are very different from the case of nova outbursts,
in which most envelope mass is blown off. 
It is because the difference is in the nuclear energy release
per unit mass between hydrogen-burning and helium-burning: 
hydrogen burning produces energy 
ten times larger than that of helium burning, i.e.,
only a tenth of the hydrogen-rich envelope mass is enough to blow 
the entire envelope off.  

Figure 4 shows the accumulation efficiency defined 
by the ratio of the processed matter remaining after one cycle of helium 
shell flash to the ignition mass.  The amount of matter lost by the 
wind in the decay phase is calculated from the rate of 
wind mass-loss and nuclear burning in Figure 3. 
This accumulation ratio is well fitted by the solid curve expressed by 
\begin{equation}
\eta_{\rm He} = \left\{ 
      \begin{array}{@{\,}ll}
1,  
& \mbox{~for~} 
-5.9 \le \log \dot M \ltsim -5 \cr
          -0.175 \left( \log \dot M + 5.35 \right)^2 + 1.05,
& \mbox{~for~}
-7.3 < \log \dot M < -5.9 
       \end{array}
    \right.
\label{helium_accumulation}
\end{equation}
where $\dot M$ is in units of $M_\odot$ yr$^{-1}$.

\section{DISCUSSIONS}
Cassisi et al (1998, CIT98) have followed the evolution of 
hydrogen-accreting white dwarfs.
These WDs are undergoing many cycles of hydrogen shell 
flashes until a strong helium shell flash develops 
to expand the envelope to a red-giant dimension. 
Based on these calculations with the old opacity, 
they concluded that a hydrogen-accreting white 
dwarf will not become an SN Ia, because almost all 
of the helium envelope is lost from the system by a Roche lobe overflow 
or a common envelope ejection.
CIT98 further denied the possibility of HKN96's scenario. 

In what follows, we criticize CIT98's argument and support the growth 
of C+O WDs as expected in HKN96's and Li \& van den Heuvel's (1997) models.
CIT98 calculated only much 
less massive white dwarfs of $0.516 M_\odot$ and $0.8 M_{\odot}$ 
to criticize $\sim 1.3 M_{\odot}$ model of HKN96. 
The trends of the envelope expansion and the mass accumulation 
depend on the WD mass because of the large difference 
in the surface gravity; 
nuclear energy release in helium burning is comparable to 
the gravitational potential energy for massive WDs, i.e., 
the ratio of nuclear/gravity is $\sim 1.0$ for a $1.3 M_{\odot}$ WD, 
whereas it is as large as $3.9$ 
for a $0.8 M_{\odot}$ WD, and $8.9$ for a $0.5 M_{\odot}$ WD. 
For massive WDs, therefore, a large part of helium is consumed 
to produce nuclear energy before the envelope expands against 
the gravity. 
It reduces the helium content of the envelope.  
This low helium content in the envelope increases 
the burning speed in mass coordinates 
as understood from 
$\dot M_{\rm nuc} \approx L_{\rm Edd}/Y\varepsilon_{\rm He}$,
where $\varepsilon_{\rm He}$ is the nuclear energy release 
per unit mass for helium burning.
Thus, CIT98's criticism is inappropriate in the sense that 
their results for less massive WDs cannot be directly applied 
to very massive WDs.

Once the optically thick wind occurs, 
a Roche lobe overflow or a common envelope ejection discussed
in CIT98 becomes physically meaningless, because
the wind velocity $\sim 1000$ km s$^{-1}$ is 
much faster than the orbital velocity of the companion
($\ltsim 300$ km s$^{-1}$) of typical SN Ia progenitors
(WD+MS or WD+RG model). 
The wind goes away quickly without any effective acceleration
by the companion.   After the wind stops, 
the photospheric radius ($\ltsim 0.1 R_\odot$)
is much smaller than the Roche lobe ($\gtsim 1 R_\odot$).
Therefore, neither a Roche lobe overflow nor
a common envelope ejection works in our binary system.

We have ignored hydrogen-rich envelope above the helium layer.
An accreting 1.3 $M_{\odot}$ WD has a hydrogen-rich envelope of 
$2 \times 10^{-7} - 3 \times 10^{-6} M_{\odot}$ above the helium layer, 
depending on the mass transfer rate, chemical composition, 
and the other details of binary models. This is much smaller 
($\ltsim 0.01$) than 
the mass of the helium layer at 
ignition and we have neglected its contribution 
to the mass accumulation efficiency.  
The hydrogen-rich envelope will be quickly blown off at a very 
early stage of helium shell flashes. 

Helium shell flashes may be observed as transient super-soft X-ray  
sources.  For high mass accretion rates as expected in SN Ia progenitors, 
shell flashes are so weak that the surface temperature never 
decrease to less than $\sim 1 \times 10^5$ K. 
Therefore, it will be a bright super-soft X-ray phenomenon 
with a faint optical counterpart. 
For a $1.3 M_\odot$ WD, the bright SSS stage lasts 
$\sim 1-3$ yrs with recurrence 
period of a few tens to several tens of years. 
For more massive WDs just before explosions, 
it lasts several
months with the recurrence period of several years.

\acknowledgments
     This research has been supported in part by the Grant-in-Aid for
Scientific Research (08640321, 09640325) of the Japanese Ministry 
of Education, Science, Culture, and Sports.

\begin{figure}
\plotone{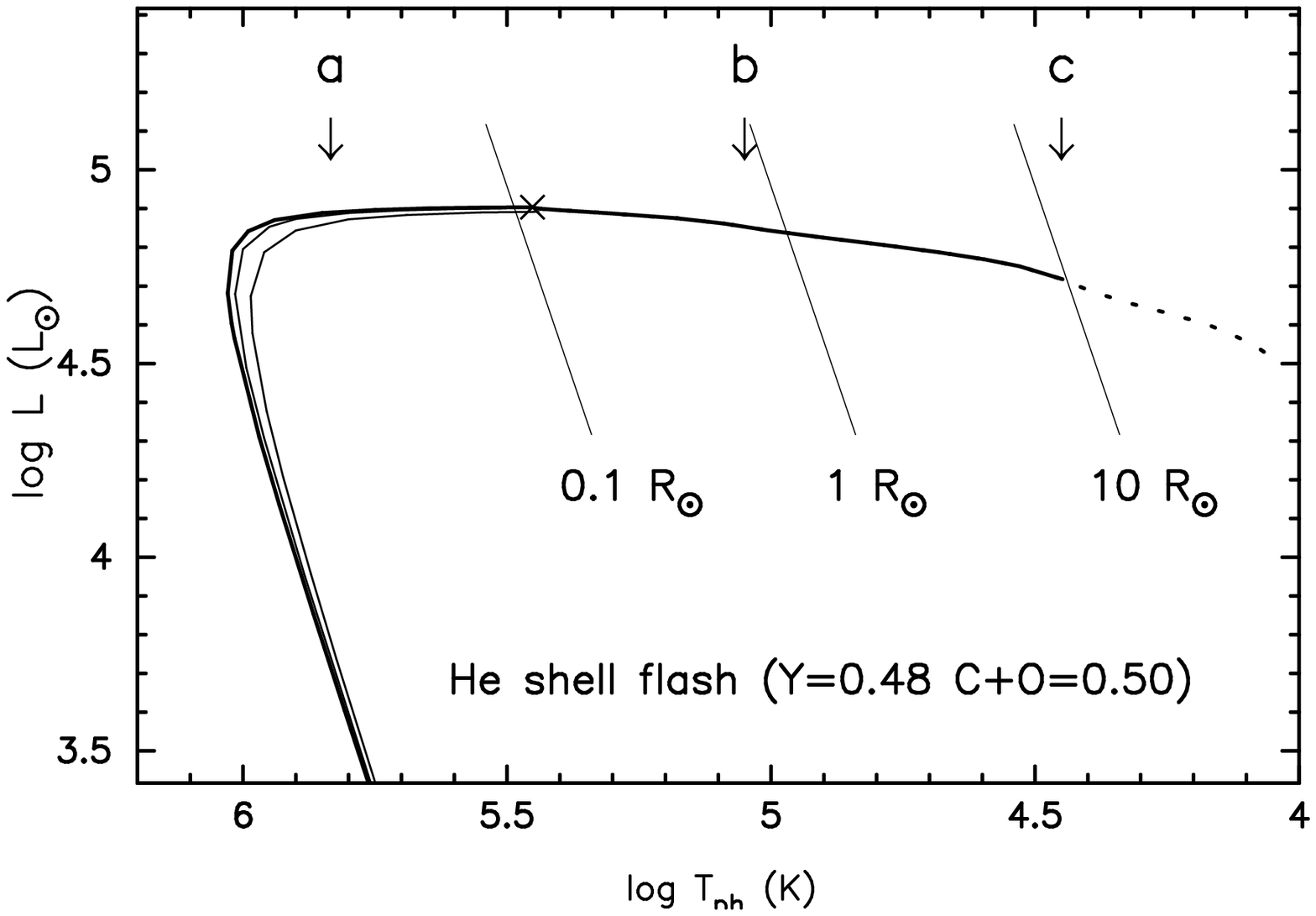}
\caption{
Evolutionary tracks of one cycle of helium shell flash 
on a 1.3 M$_{\odot}$ white dwarf. Two tracks of the rising phase 
are shown by thin solid curves: the ignition mass is 
(b) $\Delta M_{\rm ig}= 8.7 \times 10^{-5} M_\odot$ (upper), and
(c) $3.5 \times 10^{-4} M_\odot$ (lower). 
The maximum expansion of the photosphere 
is indicated by an arrow for three ignition masses of 
(a) $\Delta M_{\rm ig}= 4.8 \times 10^{-5} M_\odot$, 
(b) $\Delta M_{\rm ig}= 8.7 \times 10^{-5} M_\odot$, and 
(c) $\Delta M_{\rm ig}= 3.5 \times 10^{-4} M_\odot$. 
The decay phases for these three ignition masses are merged into
one line (thick solid).
On the decay phase, the optically thick wind ceases 
at the point marked by the cross.
The decay phase is extended to $\log T = 4.07$ (short-dashed)
for a strong helium shell flash 
with a large initial envelope mass
$\Delta M_{\rm ig}= 7.7 \times 10^{-4} M_\odot$. 
 \label{fig1}}
\end{figure}

\begin{figure}
\plotone{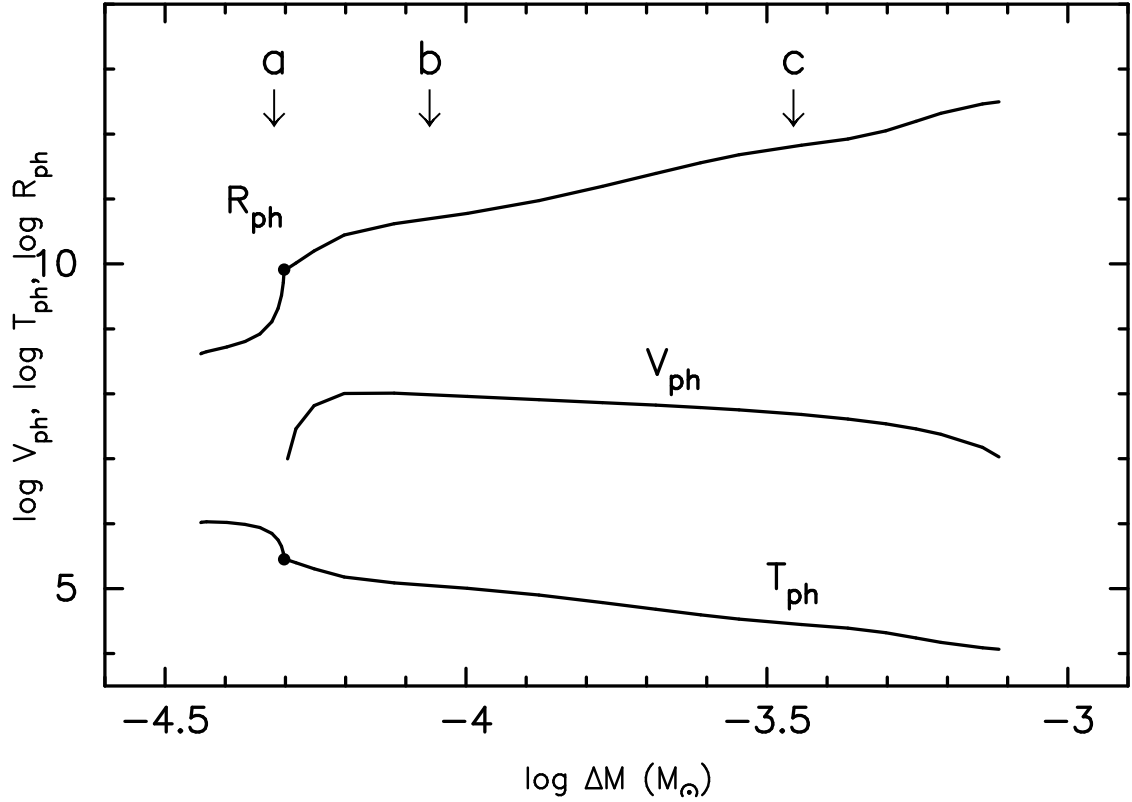}
\caption{
Photospheric radius $R_{\rm ph}$ (cm), temperature 
$T_{\rm ph}$ (K), and wind velocity 
$V_{\rm ph}$ (cm s$^{-1}$), 
are plotted against the envelope mass $\Delta M$ 
for the decay phase of helium shell flashes. Time runs from right to left.
The wind mass-loss stops at $\Delta M= 5.0 \times 10^{-5} M_\odot$  
denoted by the dots. 
The three ignition masses shown in Fig. 1 are indicated by
the arrows.
\label{fig2}  }
\end{figure}

\begin{figure}
\plotone{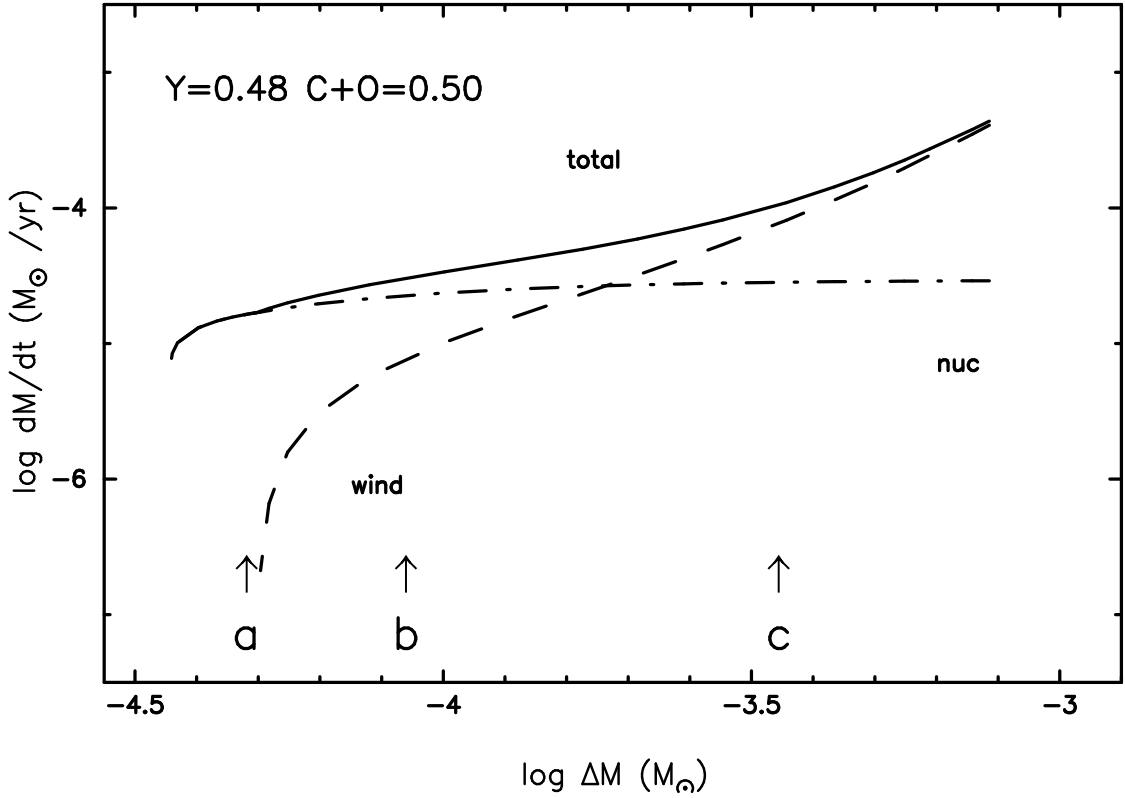}
\caption{
Wind mass-loss rate $\dot M_{\rm wind}$ (dashed), 
the mass decreasing rate owing to helium
nuclear burning $\dot M_{\rm nuc}$ (dot-dashed), 
and the total envelope mass decreasing rate 
$\dot M_{\rm wind}+ \dot M_{\rm nuc}$ (solid) 
are plotted against the envelope mass $\Delta M$.  
The ignition mass of 
three cases are shown by the arrows. Time runs from right to left.
\label{fig3}  }
\end{figure}

\begin{figure}
\plotone{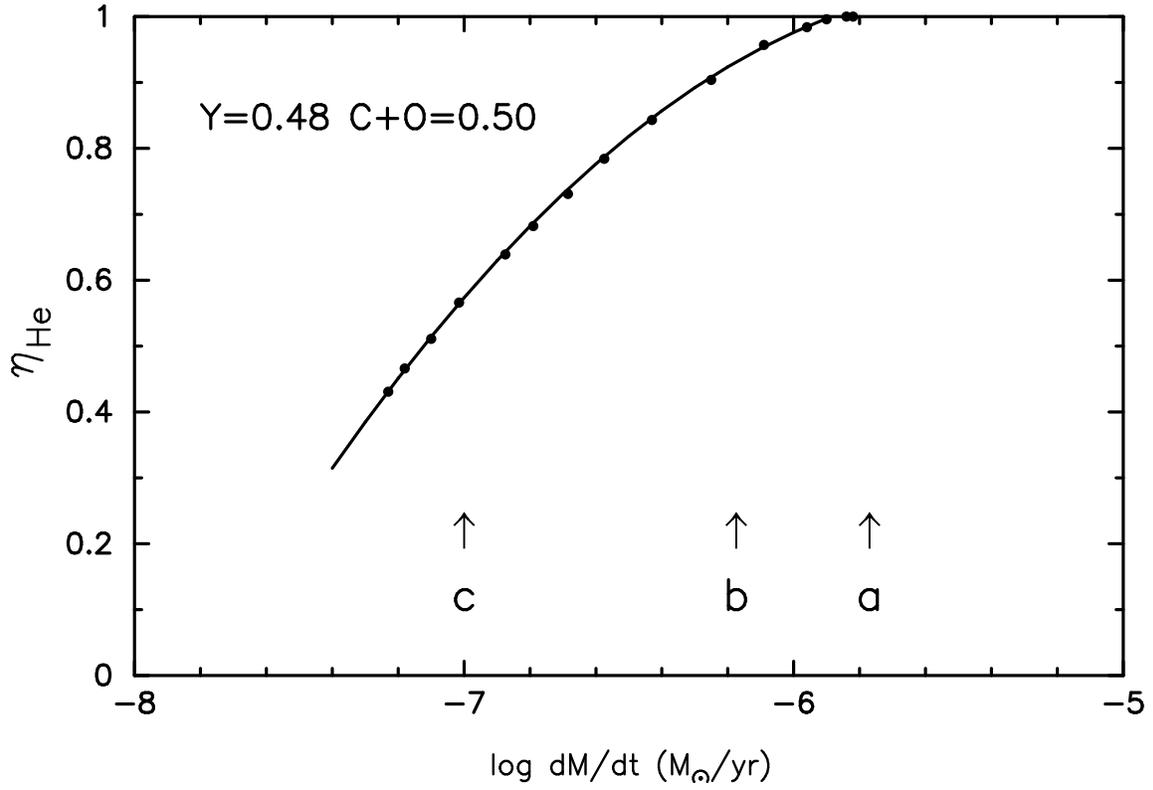}
\caption{
Mass accumulation efficiency $\eta_{\rm He}$, the ratio of
the mass accreted after one cycle of helium shell flash to the ignition
mass, is plotted against the helium accretion rate (filled circles).
The solid curve is an empirical formula given by equation (1).
The three mass accretion rates are indicated by the arrows, which
are corresponding to the three ignition masses shown in Fig. 1.
\label{fig4}}
\end{figure}  


\end{document}